\documentclass[12pt]{iopart}

\usepackage{iopams,graphicx,amssymb}
\eqnobysec

\begin{document}

\title[Critical behaviour of the stirred dynamical Potts model]
{Critical behaviour of the randomly stirred dynamical Potts model:
Novel universality class and effects of compressibility}

\author{ N V Antonov and A S Kapustin}

\address{Department of Theoretical Physics, St.~Petersburg State University
\\ Uljanovskaja 1, St.~Petersburg, Petrodvorez, 198504 Russia}

\ead{nikolai.antonov@pobox.spbu.ru, dyens@mail.ru}

\begin{abstract}
Critical behaviour of a nearly critical system, subjected to vivid turbulent
mixing, is studied by means of the field theoretic renormalization group.
Namely, relaxational stochastic dynamics of a non-conserved order parameter
of the Ashkin-Teller-Potts model, coupled to a random velocity field with
prescribed statistics, is considered. The mixing is modelled by Kraichnan's
rapid-change ensemble: time-decorrelated Gaussian velocity field with the
power-like spectrum $\propto k^{-d-\xi}$. It is shown that, depending on the
symmetry group of the underlying Potts model, the degree of compressibility
and the relation between the
exponent $\xi$ and the space dimension $d$, the system exhibits various types
of infrared (long-time, large-scale) scaling behaviour, associated with four
different infrared attractors of the renormalization group
equations. In addition to known asymptotic regimes (equilibrium dynamics of
the Potts model and the passively advected scalar field), existence of a
new, strongly non-equilibrium type of critical behaviour is established.
That ``full-scale'' regime corresponds to the novel type of critical
behaviour  (universality class), where the self-interaction of the order
parameter and the turbulent mixing are equally important. The corresponding
critical dimensions depend on $d$, $\xi$, the symmetry group and the degree
of compressibility. The dimensions and the regions of stability for all the
regimes are calculated in the leading order of the double expansion in two
parameters $\xi$ and $\varepsilon=6-d$. Special attention is paid to the
effects of compressibility of the fluid, because they lead to nontrivial
qualitative crossover phenomena.
\end{abstract}


\maketitle

\section{Introduction} \label{sec:Intro}

Numerous complex systems, involving ``infinitely'' many degrees of
freedom, demonstrate very interesting behaviour in the vicinity of
their critical points. The typical feature of such systems is the
divergence of the correlation length, although the underlying
microscopic interactions are short-range ones. As a result, the
behaviour of such systems ``decouples'' from their specific
physical nature (fluids, magnets, superconductors {\it etc}.) and
becomes highly universal: their thermodynamic and correlation
functions exhibit scaling (self-similar) behaviour and depend only
on ``global'' characteristics of the system (like its symmetry and
dimension). Thus one can speak about the theory of critical {\it
state} (or critical {\it behaviour}), in contradistinction with
the numerous special theories of diverse critical {\it phenomena}
(of course, very interesting and important on their own).

The adequate theory of critical state (both qualitative and quantitative)
is based on the field theoretic renormalization group (RG); see the
monographs \cite{Zinn,Book3} for the detailed presentation and the
bibliography. In the RG approach, possible types of critical behaviour
(universality classes) are associated with infrared (IR) attractive fixed
points of certain renormalizable field theoretic models.

Most typical equilibrium phase transitions belong to the universality class
of the $O(n)$-symmetric $\psi^{4}$ model of an $n$-component scalar order
parameter $\psi$. Universal characteristics of the critical behaviour
(critical exponents, amplitude ratios, scaling functions) depend only on
$n$ and on the spatial dimension $d$ and can be calculated within systematic
perturbative schemes, in particular, in the form of expansions in
$\varepsilon=4-d$, the deviation of the space dimension from its upper
critical value $d=4$.

More exotic, but still actual and important example is provided by the
Ashkin--Teller--Potts (or simply Potts) class of models
\cite{ATP}--\cite{Bonfim}. In rough words, such a model describes a certain
system which locally has $n$ states, while the energy of any given global
configuration depends on whether the pairs of neighboring sites are in the
same state or not. In the continuous formulation, Potts-type models are
conveniently represented by the effective Hamiltonian for the $n$-component
order parameter with a trilinear interaction term, invariant under the
$n$-dimensional hypertetrahedron symmetry group \cite{Golner}--\cite{Bonfim}.

The Potts-type models have numerous physical applications: solids and
magnetic materials with nontrivial symmetry, spin glasses,
nematic-to-isotropic transitions in the liquid crystals, bond percolation
problem, random resistor networks and many others, see the papers
\cite{Golner}--\cite{Bonfim} and the references therein. The Potts model
(especially its paradigmatic special case $n=d=2$, the famous two-dimensional
Ising model) has long become a source of inspiration for the new physical and
mathematical ideas like integrability, conformal invariance, discrete
holomorphisity {\it etc.}; for a recent discussion, see the papers
\cite{VVV}, materials of the meeting \cite{CIDHI} and references therein.

The problem of the nature of the phase transition in the Potts model has
a long and entangled history; see {\it e.g.} the discussion in \cite{Amit}.
According to naively applied Landau's phenomenological approach,
existence of a triple term excludes the possibility of the second-order
transition. On the contrary, exact two-dimensional results, numerical
simulations and RG analysis suggest that for $n$ small enough, the phase
transition in the Potts models is of the second order. We will not try to
shed new light onto this interesting but difficult question: it should first
be solved for the original {\it static} problem. To proceed, in this paper
we accept the point of view that the existence of an IR attractive fixed
point of the RG equations implies existence of a certain scaling
(self-similar) IR asymptotic regime and, therefore, the existence of some
kind of critical state.

It has long been realized that the behaviour of a real system near its
critical point is extremely sensitive to external disturbances, gravity,
geometry of the experimental setup, inclusion of impurities and so on;
see the nice book \cite{Ivanov} for the general discussion and references.
The ideal critical behaviour of a pure equilibrium infinite system can be
distorted by finite-size effects, limited accuracy of measuring the
temperature, finite time of evolution (ageing) {\it etc}.
What is more, some disturbances (randomly distributed impurities in
magnets and turbulent mixing of fluid systems) can change the very type
of the phase transition (first-order to second order one, and vice versa)
and produce novel types of critical behaviour (universality classes) with
rich and rather unexpected properties.

Investigation of the effects of various kinds of deterministic or chaotic
flows on the behaviour of the critical fluids (like liquid crystals or
binary mixtures) has shown that the flow can destroy the usual critical
behaviour: it can change to the mean-field behaviour or, under some
conditions, to a more complex behaviour described by new non-equilibrium
universality classes \cite{Beysens}--\cite{Anipoz}.

In this paper we apply the field theoretic RG to study the effects of
turbulent mixing on the critical behaviour of the (generalized) Potts model.
Special attention will be paid to compressibility of the fluid, because it
can lead to interesting crossover phenomena.

Bearing in mind applications to liquid crystals or percolation in moving
media, we consider a purely relaxational stochastic dynamics of a
non-conserved order parameter, coupled to a random velocity field.

The latter is modelled by the Kraichnan's rapid-change ensemble:
time-decorrelated Gaussian field with the pair correlation function of the
form $\langle vv\rangle \propto \delta(t-t') \, k^{-d-\xi}$, where $k$ is
the wave number and $0<\xi<2$ is a free parameter with the most realistic
(``Kolmogorov'') value $\xi=4/3$. Some time ago, the models involving passive
scalar fields advected by such ``synthetic'' velocity ensembles attracted a
great deal of attention because of the insight they offer into the origin of
intermittency and anomalous scaling in the real fluid turbulence; see the
review paper \cite{FGV} and references therein. The RG approach to the
problem of passive advection is reviewed in \cite{JphysA}.

In the context of our study it is especially important that Kraichnan's
ensemble allows one to easily model compressibility of the fluid, which
appears rather difficult if the velocity is modelled by dynamical
equations; cf.~\cite{AK}.

The outline of the paper is the following. In section~\ref{sec:QFT} we give
the detailed description of the model and its field theoretic formulation.
Following \cite{Bonfim}, we consider the generalized model with a certain
symmetry group ${\cal G}$; the plain Potts model corresponds to the group
$Z_{n}$. In section~\ref{sec:Reno} we analyze the ultraviolet (UV)
divergences and show that the model is multiplicatively renormalizable.
Thus the RG equations are derived in the standard fashion; see
section~\ref{sec:RGE}. The corresponding RG functions ($\beta$ functions
and anomalous dimensions $\gamma$) are calculated in the one-loop
approximation.

In section~\ref{sec:FPS} we show that the model reveals four different
types of critical behaviour, associated with four possible fixed points
(strictly speaking, attractors of a more general form) of the corresponding
RG equations. Three regimes correspond to known situations: Gaussian or free
field theory, passively advected scalar field without self-interaction
(the nonlinearity in the original stochastic equation is irrelevant), and
the original Potts model without mixing. The most interesting fourth regime
corresponds to a novel non-equilibrium universality class, formed by the
interplay between the self-interaction and turbulent mixing.

A given critical regime can realize if the corresponding fixed point is
admissible from the physics viewpoints: it should be IR attractive and lie
in the physical range of model parameters. In our problem, admissibility
of the fixed points depends on the spatial dimension $d$, exponent $\xi$,
symmetry group ${\cal G}$ and the degree of compressibility. The regions of
admissibility form a pattern in the space of model parameters, much more
complicated in comparison with the $\psi^{4}$ model in the same velocity
ensemble \cite{AK} or the Potts model in a strongly anisotropic shear flow
\cite{Anipoz}. As a result, interesting crossover phenomena occur as the
degree of compressibility is varied. These issues are discussed in great
detail in the same section~\ref{sec:FPS}.

In general, the critical dimensions in our problem are calculated as double
series in $\varepsilon=6-d$ and $\xi$. In section~\ref{sec:DimeNS} we present
the explicit first-order (one-loop) expressions for the critical dimensions
of the basic fields and parameters. Section~\ref{sec:Conc}  is reserved for
a brief conclusion.

\section{Description of the model. Field theoretic formulation}
\label{sec:QFT}

Relaxational dynamics of a non-conserved $n$-component order parameter
$\psi_a(x)$ with $x\equiv\{t,\textbf{x}\}$ is described by the
Langevin-type stochastic differential equation
\begin{eqnarray}
\partial_{t} \psi_a(x) =-\lambda_0 \, \frac{\delta {\cal H}(\psi)}{
\delta \psi_a({\bf x})} \Bigg{|}_{{\bf x} \to x} + \zeta_a(x),
\label{eq1}
\end{eqnarray}
where $\partial_{t} = \partial/ \partial {t}$, $\lambda_0>0$ is the
(constant) kinetic coefficient  and $\zeta_a(x)$ is a Gaussian random noise
with zero average and the pair correlation function
\begin{eqnarray}
\langle \zeta_a(x)\zeta_b(x') \rangle = 2\lambda_0\,\delta_{ab} \,
\delta(t-t')\, \delta^{(d)}(\bf x-\bf x') ,
\label{forceD}
\end{eqnarray}
where $d$ is the dimension of the $\bf x$ space. Here and below, the bare
(unrenormalized) parameters are marked by the subscript ``0.'' Their
renormalized counterparts (without the subscript) will appear later on.
The normalization in (\ref{forceD})  is chosen such that the steady-state
equal-time correlation functions of the stochastic problem are given by the
Boltzmann weight $\exp\{-{\cal H}(\psi)\}$.

Near the critical point, the static Hamiltonian ${\cal H}(\psi)$ of
the Potts model is taken in the form \cite{Golner,Zia,Priest}
\begin{eqnarray}
{\cal H}(\psi) &=&  \int d{\bf x} \left\{- \frac{1}{2}\,
 \psi_a({\bf x})\partial^{2} \psi_a({\bf x}) + \frac{\tau_{0}}{2}\,
\psi_a({\bf x})\psi_a({\bf x}) + \right.
\nonumber \\
&+& \left. \frac{g_{0}}{3!}R_{abc} \,
\psi_a({\bf x})\psi_b({\bf x})\psi_c({\bf x}) \right\},
\label{LG}
\end{eqnarray}
where $\partial_{i} = \partial/ \partial x_{i}$ is the spatial derivative,
$\partial^{2} = \partial_{i}\partial_{i}$ is the Laplace operator,
$\tau_{0} \propto (T-T_{c})$ measures deviation of the temperature
(or its analog) from the critical value and $g_0$ is the coupling constant.
Summations over repeated indices are always implied
($a,b,c=1,\dots,n$ and $i=1,\dots,d$).
After taking the functional derivative
$\delta {\cal H}(\psi) / \psi({\bf x})$ one has to replace
$\psi({\bf x})$ in (\ref{eq1}) by the time-dependent field $\psi(x)$.

Following \cite{Bonfim}, we consider the generalized case of certain symmetry
group ${\cal G}$, which has the only irreducible invariant third-rank tensor
$R_{abc}$; without loss of generality it is assumed to be symmetric.
In the one-loop approximation we will only need to know the coefficients
$R_{1,2}$ in the contractions
\begin{eqnarray}
R_{abc}R_{abe}=R_1 \delta_{ce}, \quad
R_{aec}R_{chb}R_{b\!f\!a}=R_2 R_{eh\!f}.
\label{rr}
\end{eqnarray}

In the original Potts model ${\cal G}=Z_{n}$, the symmetry group of the
hypertetrahedron in the $n$-dimensional space. Then it is convenient to
express the tensor $R_{abc}$ in terms of the set of $(n+1)$ vectors
$e^{\alpha}$ which define its vertices \cite{Golner,Zia}:
\[ R_{abc}=\sum_{\alpha}e^{\alpha}_ae^{\alpha}_be^{\alpha}_c, \]
where the $e^{\alpha}_a$ satisfy
\begin{eqnarray}
\sum_{\alpha=1}^{n+1}e^{\alpha}_a=0, \quad
\sum_{\alpha=1}^{n+1}e^{\alpha}_ae^{\alpha}_b=(n+1)\delta_{ab}, \quad
\sum_{a=1}^{n} e^{\alpha}_ae^{\beta}_a=(n+1)\delta^{\alpha\beta}-1.
\label{tensor2}
\end{eqnarray}
For the coefficients in (\ref{rr}) the relations (\ref{tensor2}) give
\begin{eqnarray}
R_1=(n+1)^2(n-1), \quad  R_2=(n+1)^2(n-2).
\label{contract}
\end{eqnarray}

The stochastic problem (\ref{eq1}), (\ref{forceD}) can be
reformulated as the field theoretic model of the doubled set of fields
$\Phi = \{\psi,\psi^{\dag}\}$ with the action functional
\begin{eqnarray}
{\cal S} (\psi,\psi^{\dag}) =
\lambda_{0} \psi_{a}^{\dagger}\psi_{a}^{\dagger} \!+\!
\psi_{a}^{\dag} \left(-\partial_{t}+\lambda_{0} \partial^{2} \!-\!
\lambda_{0}\tau_{0}\right) \psi_{a} \!-\!
g_{0}\lambda_{0} R_{abc} \psi^{\dag}_{a}\psi_{b}\psi_{c}/2.
\label{action}
\end{eqnarray}
Here $\psi^{\dag}=\psi^{\dag}(t,{\bf x})$ is an auxiliary
``response field'' and all the needed integrations over the arguments of the
fields and summations over the repeated indices are implied, for example
\[  \psi_{a}^{\dag}\partial_{t}\psi_{a} = \sum_{a=1}^{n}
\int dt \int d{\bf x}
\psi_{a}^{\dag}(t,{\bf x})\partial_{t}\psi_{a}(t,{\bf x}). \]

This formulation means that the statistical averages of the random quantities
in the original stochastic problem can be represented by functional integrals
over the full set of fields with the weight $\exp {\cal S}(\Phi)$, and can
therefore be interpreted as the Green functions of the field theoretic model
with the action (\ref{action}).

The model (\ref{action}) corresponds to the standard Feynman
diagrammatic technique with two bare propagators
$\langle \psi \psi^{\dag} \rangle_{0}$, $\langle \psi \psi \rangle_{0}$
and the trilinear vertex $\sim \psi^{\dagger}\psi^2$. In the
frequency-momentum ($\omega$--${\bf k}$) representation the propagators
have the forms
\begin{eqnarray}
\langle \psi_{a} \psi_{b}^{\dag} \rangle_{0} (\omega,{\bf k}) =
\frac{\delta_{ab}}{-{\rm i}\omega+\lambda_{0} \left(k^{2}+\tau_{0}\right)},
\nonumber \\
\langle \psi_{a} \psi_{b} \rangle_{0}(\omega,{\bf k}) =
\frac{2\lambda_{0}\delta_{ab}}{\omega^2+\lambda_{0}^2
\left(k^{2}+\tau_{0}\right)^2},
\label{lines}
\end{eqnarray}
where $k=|{\bf k}|$ is the wave number.

The Galilean invariant coupling with the velocity field
${\bf v}= \{ v_{i}(t, {\bf x}) \}$ for the compressible fluid
($\partial _i v_{i} \ne 0$) is introduced by the replacement
\begin{eqnarray}
\partial_{t}\psi \to \partial_{t}\psi + a_{0}\, \partial_{i}(v_{i}\psi) +
(a_{0}-1) (v_{i} \partial_{i})\psi =
\nabla_{t} \psi + a_{0}(\partial_{i}v_{i}) \psi
\label{nabla}
\end{eqnarray}
in (\ref{eq1}). Here $\nabla_{t} \equiv \partial_{t} + v_{i} \partial_{i}$
is the Galilean covariant (Lagrangian) derivative and $a_{0}$ is an
arbitrary parameter. For the linear advection-diffusion equation, the
choice $a_{0}=1$ corresponds to the conserved quantity $\psi$ ({\it e.g.}
density of an impurity) while $a_{0}=0$ is referred to as ``tracer''
(concentration of the impurity or the temperature of the fluid; in this
case, the conserved quantity is $\psi^{\dag}$). In the presence of a
nonlinearity in (\ref{eq1}), it is necessary to keep all the terms in
(\ref{nabla}) in order to ensure multiplicative renormalizability \cite{AK}.

In the real problem, the field ${\bf v}(t,{\bf x})$ is governed by the
Navier--Stokes equation. Here we employ a simplified rapid-change model,
in which the velocity obeys a Gaussian statistics with zero average and
the prescribed correlation function
\begin{eqnarray}
\langle v_{i}(t, {\bf x}) v_{j}(t',{\bf x'})\rangle =  \delta(t-t')\,
D_{ij}({\bf r}), \quad {\bf r} = {\bf x}-{\bf x'}
\label{white}
\end{eqnarray}
with
\begin{eqnarray}
D_{ij}({\bf r}) = D_{0}\, \int_{k>m} \frac{d{\bf k}}{(2\pi)^{d}} \,
\frac{1}{k^{d+\xi}}\,
\left\{ P_{ij}({\bf k})+\alpha Q_{ij}({\bf k}) \right\}\,
\exp ({\rm i} {\bf kr} ).
\label{Kraich}
\end{eqnarray}
Here $P_{ij}({\bf k}) = \delta_{ij} - k_i k_j / k^2$ and
$Q_{ij}({\bf k})=k_i k_j/k^2$ are the transverse and the longitudinal
projectors, respectively,
$D_{0}>0$ is an amplitude factor and $\alpha\ge0$ is an arbitrary
parameter which measures the degree of compressibility.
The case $\alpha=0$ corresponds to the incompressible fluid
($\partial _i v_{i}=0$), while the limit $\alpha \to\infty$ at fixed
$\alpha D_{0}$ corresponds to the purely potential velocity field.
The exponent $0<\xi<2$ is a free parameter which can be viewed as a kind
of H\"{o}lder exponent, which measures ``roughness'' of the velocity field;
its ``Kolmogorov'' value is $\xi=4/3$, while the ``Batchelor'' limit
$\xi\to2$ corresponds to smooth velocity. The cutoff in the
integral (\ref{Kraich}) from below at $k=m$, where $m\equiv 1/{\cal L}$ is
the reciprocal of the integral turbulence scale ${\cal L}$, provides IR
regularization. Its precise form is inessential; the sharp cutoff is the
simplest choice from calculational viewpoints.

The action functional for the full set of fields $\Phi = \left\{
\psi,\psi^{\dag},{\bf v} \right\}$ becomes
\begin{eqnarray}
{\cal S}(\Phi) &=& \lambda_{0} \psi^{\dag}_{a}\psi^{\dag}_{a}+
\psi^{\dag}_{a} \left\{ -\nabla_{t} +
\lambda_{0}\left( \partial^{2}- \tau_{0}\right)
- a_{0} (\partial_{i}v_{i}) \right\} \psi_{a} +
\nonumber \\
 &-&  \frac{g_{0}\lambda_{0}}{2}R_{abc} \psi^{\dag}_{a}
\psi_{b}\psi_{c} +  {\cal S}_{v}({\bf v}).
\label{Action}
\end{eqnarray}
It is obtained from (\ref{action}) by the substitution
(\ref{nabla}) and adding the term corresponding to the Gaussian averaging
over the velocity field with the correlator (\ref{white}), (\ref{Kraich}):
\begin{eqnarray}
{\cal S}_{v}({\bf v}) = -\frac{1}{2} \int dt \int d{\bf x} \int d{\bf x'}
v_{i} (t,{\bf x}) D_{ij}^{-1}({\bf r}) v_{j} (t, {\bf x'}),
\label{Sv}
\end{eqnarray}
where $D_{ij}^{-1}({\bf r}) \propto D_{0}^{-1} r^{-2d-\xi}$ with
$r=|{\bf r}|$ is the kernel of the linear operation inverse to
$D_{ij}({\bf r})$ from (\ref{Kraich}).

In addition to (\ref{lines}), the diagrammatic technique for the
full-scale model involves the velocity propagator $\langle vv \rangle_{0}$
defined by the relations (\ref{white}), (\ref{Kraich}) and the
new vertex
\begin{equation}
\psi_{a}^{\dag} v_{i} V_{i,ab} \psi_{b} \equiv
- \psi_{a}^{\dag}\left\{
(v_{i}\partial_{i}) \psi_{a} + a_{0}(\partial_{i}v_{i}) \psi_{a} \right\}
\label{Vertex}
\end{equation}
with the vertex factor
\begin{equation}
V_{i,ab} = - {\rm i} \delta_{ab} (k_{i}+ a_{0} q_{i}),
\label{VertexF}
\end{equation}
where $q_{i}$ is the momentum argument of $v_{i}$ and
$k_{i}$ is the momentum of the field $\psi$.

As usual for a trilinear interaction, the actual expansion parameter
in the model (\ref{action}) is $u_{0}\equiv g_{0}^2$
rather than $g_{0}$ itself. Thus for the full model (\ref{Action}) the part
of the coupling constants (expansion parameters in the ordinary perturbation
theory) is played by the three parameters
\begin{equation}
u_{0}=g_{0}^2  \sim \Lambda^{6-d}, \quad w_{0} =
D_{0}/\lambda_{0} \sim \Lambda^{\xi}, \quad w_{0}a_{0} \sim
\Lambda^{\xi}.
\label{charges}
\end{equation}
The last relations follow from the dimensionality considerations (more
precisely, see the next section) and set in the characteristic UV momentum
scale $\Lambda$. The first relation in (\ref{charges}) shows
that the interaction $\psi^{\dagger}\psi^{2}$ becomes logarithmic
(the corresponding coupling constant $u_{0}$ becomes dimensionless) for
$d=6$. Thus for the single-charge problem (\ref{action}) the value
$d=6$ is the upper critical dimension, and the deviation $\varepsilon=6-d$
plays the role of the expansion parameter in the RG approach:
the critical dimensions are nontrivial for $\varepsilon>0$ and can be
calculated in the form of series in $\varepsilon$.

The additional interaction (\ref{Vertex}) of the full model becomes
logarithmic for $\xi=0$. The parameter $\xi$ is not related to the spatial
dimension and can be varied independently. For the RG analysis of the
full-scale problem it is important that all the interactions become
logarithmic simultaneously. Otherwise, one of them would be weaker than
the other from the RG viewpoint and it would be irrelevant in the
leading-order IR behaviour. As a result, some of the scaling regimes
of the full model would be lost. In order to study all possible scaling
regimes and the crossovers between them, we need a genuine multicharge
theory, in which all the interactions are treated on equal footing. Thus
in the following we treat $\varepsilon$ and $\xi$ as small parameters of
the same order, $\varepsilon \propto \xi$. Instead of the ordinary
$\varepsilon$ expansion in the single-charge models, the coordinates
of the fixed points, critical dimensions and other quantities will be
calculated as double expansions in $\varepsilon$ and $\xi$.

\section{Canonical dimensions, UV divergences and renormalization}
\label{sec:Reno}

It is well known that the analysis of UV divergences is based on
the analysis of canonical dimensions (``power counting''); see
{\it e.g.} \cite{Zinn,Book3}. Dynamical models of the type
(\ref{action}), (\ref{Action}),
in contrast to static ones, have two independent
scales: the time scale $T$ and the length scale $L$. Thus the
canonical dimension of any quantity $F$ (a field or a parameter)
is described by two numbers, the momentum dimension $d_{F}^{k}$
and the frequency dimension $d_{F}^{\omega}$, defined
such that $[F] \sim  [L]^{-d_{F}^{k}} [T]^{-d_{F}^{\omega}}$.
By definition
\[ d_k^k=-d_{\bf x}^k=1,\ d_k^{\omega} =d_{\bf x}^{\omega }=0,\
d_{\omega }^k=d_t^k=0, \ d_{\omega }^{\omega }=-d_t^{\omega }=1, \]
while the other dimensions are found from the requirement that each
term of the action functional be dimensionless (with
respect to the momentum and frequency dimensions separately).
Then, based on $d_{F}^{k}$ and $d_{F}^{\omega}$, one can introduce
the total canonical dimension $d_{F}=d_{F}^{k}+2d_{F}^{\omega}$
(in the free theory, $\partial_{t}\propto\partial^{2}$), which
plays in the theory of renormalization of dynamical models the
same part as the conventional (momentum) dimension does in static
problems; see {\it e.g.} chap.~5 of \cite{Book3}. The canonical dimensions
for the models (\ref{Action}) are given in table~\ref{table1},
including the renormalized parameters (without subscript ``0''), which
will be introduced shortly.

\begin{table}[h!]
\caption{Canonical dimensions of the fields and parameters in the
model (\protect\ref{Action}).}
\label{table1}
\begin{center}
\begin{tabular}{|c|c|c|c|c|c|c|c|c|c|c|}
\hline
$F$  & $\psi$ & $\psi^\dagger$ & $v$ & $\lambda_0,\lambda$ & $\tau_0,\tau$ &
$m,\mu,\Lambda$ & $u_0$ & $w_0$ & $u,w,a_0,a,\alpha$ \\
\hline
$d_F^k$  & $(d-2)/2$ &  $(d+2)/2$ & $-1$ & $-2$ & $2$ & $1$ & $6-d$ &
$\xi$ & $0$ \\
\hline
$d_F^\omega$ & 0 & 0 & $1$ & $1$ & $0$ & $0$ & $0$ & $0$ & $0$ \\
\hline
$d_F$ & $(d-2)/2$ & $(d+2)/2$ & $1$ & $0$ & $2$ & $1$ & $6-d$ &
$\xi$ & $0$ \\
\hline
\end{tabular}
\end{center}
\end{table}

As already discussed in the end of the preceding section, the full
model (\ref{Action}) is logarithmic (all the coupling constants are
simultaneously dimensionless) at $d=6$ and $\xi=0$. Thus the UV
divergences in the Green functions manifest themselves as singularities
in $\varepsilon = 6-d$, $\xi$ and, in general, in their linear
combinations.

The total canonical dimension of an arbitrary 1-irreducible Green
function $\Gamma = \langle\Phi \cdots \Phi \rangle _{\rm 1-ir}$ is
given by the relation \cite{Book3}
\begin{equation}
d_{\Gamma }=d_{\Gamma }^k+2d_{\Gamma }^{\omega }= d+2-N_{\Phi
}d_{\Phi}, \label{dGamma}
\end{equation}
where $N_{\Phi}=\{N_{\psi},N_{\psi^{\dag}}, N_{v}\}$ are the
numbers of corresponding fields entering the function
$\Gamma$, and the summation over all types of the fields is
implied. The total dimension $d_{\Gamma}$ in the logarithmic theory
(that is, at $\varepsilon=\xi=0$) is the formal index of the UV
divergence $\delta_{\Gamma}=d_{\Gamma}|_{\varepsilon=\xi=0}$.
Superficial UV divergences, whose removal requires counterterms,
can appear only in the functions $\Gamma$ for which
$\delta_{\Gamma}$ is a non-negative integer.
From table~\ref{table1} and (\ref{dGamma}) we find
\begin{equation}
\delta_{\Gamma}= 8 - 2N_{\psi} - 4N_{\psi^{\dag}} - N_{v}.
\label{Inde}
\end{equation}

Dimensional analysis should be augmented by certain additional
considerations. In dynamical models of the type (\ref{Action}),
all the 1-irreducible diagrams without the fields
$\psi^{\dag}$ vanish, and it is sufficient to consider the
functions with $N_{\psi^{\dag}} \ge 1$ \cite{Book3}. Furthermore, an
important role is played by the Galilean symmetry and
the invariance  with respect to the group ${\cal G}$.
For example, the function
$\left\langle \psi^{\dag}\psi  vv\right\rangle_{\rm 1-ir}$ can be
omitted from consideration
because the corresponding counterterm $\psi^{\dag}_a\psi_a  v^2$
is forbidden by the Galilean invariance. A similar situation
occurs for the function $\left\langle \psi^{\dag} vv\right\rangle_{\rm 1-ir}$:
the possible counterterms
$\psi^{\dag}_a  (\partial_iv_k)(\partial_kv_i)$ and
$\psi^{\dag}_a  (\partial_iv_k)(\partial_i v_k)$
are forbidden by the symmetry with respect to ${\cal G}$.
In turn, owing to the symmetry with respect to ${\cal G}$, the
trilinear term in (\ref{Action}) is renormalized as a single entity.

With those restrictions, the analysis
of the expression (\ref{Inde}) shows that in our model, superficial
UV divergences can only be present in the following 1-irreducible
functions:
\begin{eqnarray}
\langle \psi^{\dag} \psi^{\dag} \rangle_{\rm 1-ir} \quad (\delta=0) \quad
{\rm with\ the\ counterterm} \quad \psi_{a}^{\dag}\psi_{a}^{\dag},
\nonumber \\
\langle \psi^{\dag} \psi \rangle_{\rm 1-ir} \quad (\delta=2) \quad
{\rm with\ the\ counterterms} \quad \psi_{a}^{\dag}\partial_{t}\psi_{a}, \
\psi_{a}^{\dag}\partial^{2}\psi_{a}, \ \psi_{a}^{\dag}\psi_{a},
\nonumber \\
\langle \psi^{\dag} \psi \psi \rangle_{\rm 1-ir} \quad (\delta=0) \quad
{\rm with\ the\ counterterm} \quad
R_{abc} \psi_{a}^{\dag} \psi_{b} \psi_{c},
\nonumber \\
\langle \psi^{\dag} \psi v \rangle_{\rm 1-ir} \quad (\delta=1) \quad
{\rm with\ the\ counterterms} \quad \psi_{a}^{\dag} (v_{i}\partial_{i})
\psi_{a}, \ \psi_{a}^{\dag} (\partial_{i} v_{i}) \psi_{a}.
\nonumber
\end{eqnarray}

All these terms are present in the action functional (\ref{Action}),
so that our model appears multiplicatively renormalizable.
The Galilean symmetry also requires that the
counterterms $\psi^{\dag}\partial_{t}\psi$ and $\psi^{\dag}
(v\partial) \psi $ enter the renormalized action only in the form
of the Lagrangian derivative $\psi^{\dag}\nabla_{t}\psi$, imposing
no restriction on the Galilean invariant term $\psi^{\dag}
(\partial v) \psi $.

We thus conclude that the renormalized action can be written in the form
\begin{eqnarray}
{\cal S}_{R} (\Phi) &=&  \psi^{\dag}_{a} \left\{ - Z_{1} \nabla_{t} +
\lambda\left( Z_{2} \partial^{2}- Z_{3}\tau\right) - a Z_{6}
(\partial_{i}v_{i}) \right\} \psi_{a}  +
 \nonumber \\
&+& \lambda Z_{5}  \psi^{\dag}_{a}\psi^{\dag}_{a} -
g\mu^{\varepsilon/2} R_{abc}
\lambda Z_{4}  \psi^{\dag}_{a}\psi_{b} \psi_{c} /2 +  {\cal S}_{v}({\bf v})
\label{ActionR}
\end{eqnarray}
with the same ${\cal S}_{v}({\bf v})$ from (\ref{Sv}).

Here $\lambda$, $\tau$, $g$ and $a$ are renormalized counterparts
of the bare parameters (with the subscripts ``0'') and $\mu$ is
the reference momentum scale (additional arbitrary parameter of the
renormalized theory). The renormalization constants $Z$ absorb
the singularities in $\varepsilon$ and $\xi$ and depend on the
dimensionless parameters $u$, $w$, $a$ and $\alpha$. Expression
(\ref{ActionR}) can be reproduced by the multiplicative
renormalization of the fields $\psi \to \psi Z_{\psi}$,
$\psi^{\dag} \to \psi^{\dag} Z_{\psi^{\dag}}$ and the parameters:
\begin{eqnarray}
g_{0} &=& g \mu^{\varepsilon/2} Z_{g}, \quad u_{0} = u
\mu^{\varepsilon} Z_{u}, \quad
w_{0} = w \mu^{\xi} Z_{w},
\nonumber \\
\lambda_{0} &=& \lambda Z_{\lambda}, \quad \tau_{0} = \tau Z_{\tau},
\quad  a_{0} = a Z_{\tau}.
\label{Multy}
\end{eqnarray}

Since the last term ${\cal S}_{v}({\bf v})$ in (\ref{ActionR}) remains
intact, the amplitude $D_{0}$ from (\ref{Kraich}) is
expressed in renormalized parameters as $D_{0} = w_{0} \lambda_{0}
= w\lambda \mu^{\xi}$, while the parameters $m$ and $\alpha$ are
not renormalized: $m_{0} = m$, $\alpha_{0} = \alpha$. Owing to the
Galilean symmetry, the both terms in the covariant derivative
$\nabla_{t}$ are renormalized with the same constant $Z_{1}$, so
that the velocity field is not renormalized, either. Hence the exact
relations
\begin{eqnarray}
Z_{w}Z_{\lambda} =1, \quad Z_{m}= Z_{\alpha} = Z_{v} =1.
\label{RenD}
\end{eqnarray}
Comparison of the expressions (\ref{Action}) and (\ref{ActionR})
gives the following relations between the renormalization
constants $Z_{1}$--$Z_{6}$ and (\ref{Multy}):
\begin{eqnarray}
Z_{1} &=& Z_{\psi} Z_{\psi^{\dagger}}, \quad Z_{2} =
Z_{1}Z_{\lambda}, \quad Z_{3} = Z_{2} Z_{\tau},
 \nonumber \\
Z_{4} &=& Z_{\lambda}Z_{g} Z_{\psi}^{2} Z_{\psi^{\dagger}}, \quad
Z_{5} = Z_{\lambda} Z_{\psi^{\dagger}}^{2}, \quad
Z_{6} = Z_{1} Z_{a}.
\label{ZZ}
\end{eqnarray}
Resolving these relations with respect to the
renormalization constants of the fields and parameters gives
\begin{eqnarray}
Z_{\lambda} &=& Z_{1}^{-1} Z_{2}, \quad
Z_{\tau} = Z_{2}^{-1} Z_{3}, \quad
Z_{a} = Z_{1}^{-1} Z_{6}, \quad
Z_{u} = Z_{1}^{-1} Z_{2}^{-3} Z_{4}^{2} Z_{5},
 \nonumber \\
Z_{\psi} &=&
Z_{1}^{1/2}Z_{2}^{1/2}Z_{5}^{-1/2}, \quad
Z_{\psi^{\dag}} = Z_{1}^{1/2}Z_{2}^{-1/2}Z_{5}^{1/2},
\label{ResoC}
\end{eqnarray}
where we have passed to the coupling constant $u=g^{2}$
with $Z_{u}=Z_{g}^{2}$.

\begin{figure}
\begin{center}
\includegraphics[width=14cm]{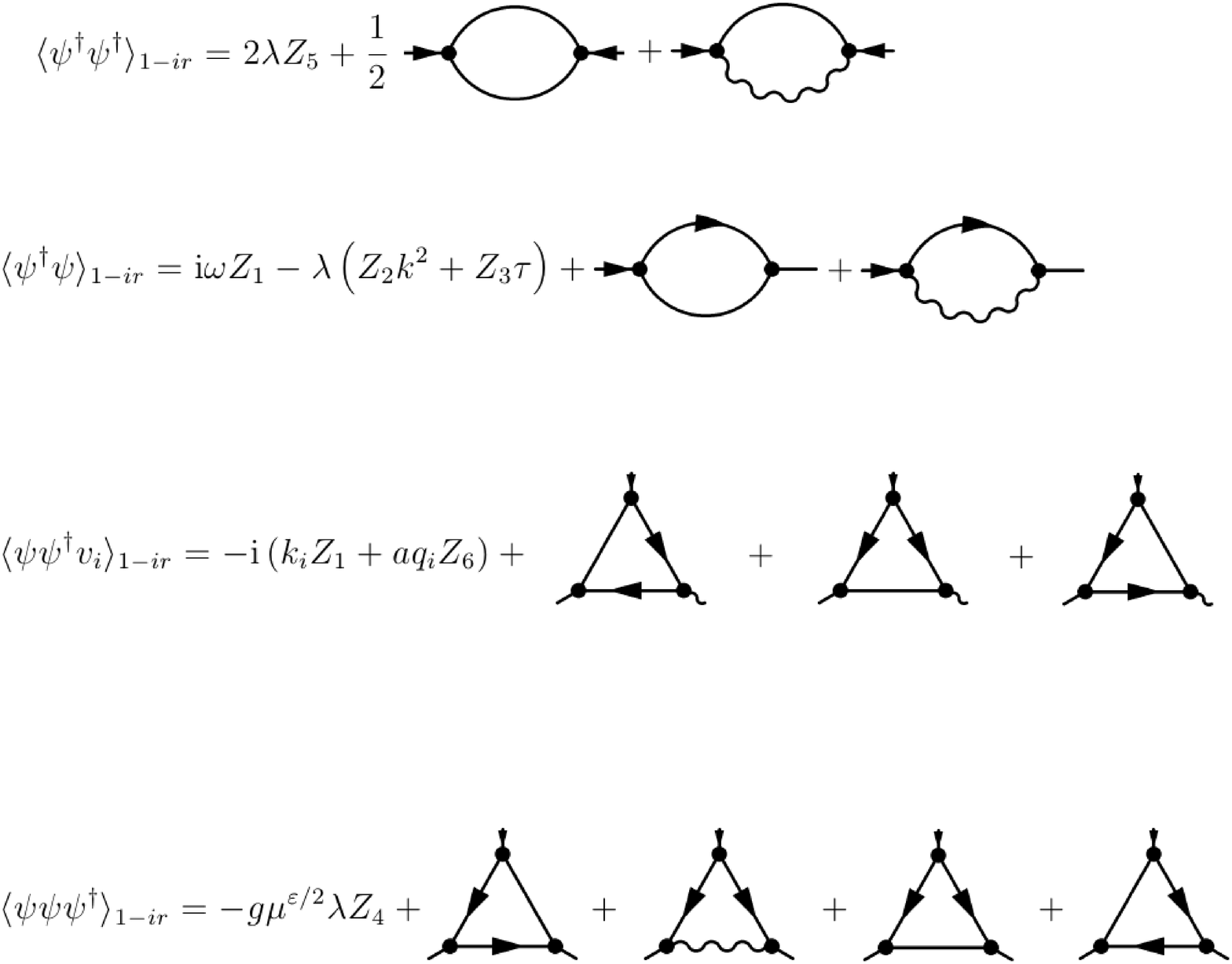}
\caption{\label{fig:diabase} Superficially divergent 1-irreducible Green
functions in the one-loop approximation.}
\end{center}
\end{figure}

The renormalization constants can be found from the requirement
that the Green functions of the renormalized model
(\ref{ActionR}), when expressed in renormalized variables, be UV
finite (in our case, finite at $\varepsilon\to0$, $\xi\to0$).
The constants $Z_{1}$--$Z_{6}$ are calculated directly from the
diagrams, then the constants in (\ref{Multy}) are found from
(\ref{ResoC}). In order to find the full set of constants, it is
sufficient to consider the 1-irreducible Green functions which
involve superficial divergences. The diagrammatic representation for
the relevant Green functions in the one-loop approximation is given
on figure~\ref{fig:diabase}.


The solid lines with arrows denote the propagator $\langle \psi \psi^{\dag}
\rangle_{0}$, the arrow pointing to the field $\psi^{\dag}$. The solid lines
without arrows correspond to the propagator $\langle \psi \psi \rangle_{0}$
and the wavy lines denote the velocity propagator $\langle vv \rangle_{0}$
defined in (\ref{white}) and (\ref{Kraich}). The external ends with incoming
arrows correspond to the fields $\psi^{\dag}$, the ends without arrows
correspond to $\psi$. The triple vertices with one wavy line correspond to
the vertex factor (\ref{VertexF}). In order to calculate the renormalization
constants, it is sufficient to consider the symmetric phase with $\tau>0$.
Then, with respect to the ``isovector'' indices, all the terms in the first
three lines on figure~1 are proportional to $\delta_{ab}$ while all the
terms in last line are proportional to $R_{abc}$.

All the diagrammatic elements should be expressed in renormalized
variables using the relations (\ref{Multy})--(\ref{ZZ}). In the
one-loop approximation, the $Z$'s in the bare terms should be
taken in the first order in $u= g^{2}$
and $w$, while in the diagrams they should simply be replaced with
unities, $Z \to 1$. Thus the passage to renormalized variables
in the diagrams is achieved by the simple substitutions
$\lambda_{0} \to \lambda$, $\tau_{0} \to \tau$, $g_{0} \to
g\mu^{\varepsilon/2}$ and $w_{0} \to w\mu^{\xi}$.

In practical calculations, we used the minimal subtraction (MS) scheme,
where all the renormalization constants have the forms
``$Z=1+\,$ only singularities in $\varepsilon$ and $\xi$,'' with
the coefficients depending on the completely dimensionless
renormalized parameters $u$, $w$, $a$ and $\alpha$.

The one-loop calculations for similar models are discussed in detail,
{\it e.g.}, in \cite{AHH}--\cite{AK}, and here we only give the results:
\begin{eqnarray}
Z_{1} &=& 1 - \frac{uR_{1}}{2\varepsilon}, \quad
Z_{2} = 1 - \frac{uR_{1}}{3\varepsilon} - \frac{w}{6\xi}(5+\alpha), \quad
Z_{3} = 1 - \frac{2R_{1}u}{\varepsilon},
\nonumber \\
Z_{4} &=& 1 - \frac{2R_{2}u}{\varepsilon} - \frac{w \alpha a^2}{\xi}, \quad
Z_{5} = 1 - \frac{uR_{1}}{2\varepsilon} - \frac{w}{\xi} \alpha (a-1)^{2},
\nonumber \\
Z_{6} &=& 1 - \frac{uR_{1}}{ 2a\varepsilon}(4a-1)
\label{Zo}
\end{eqnarray}
with $R_{1,2}$ from (\ref{rr}) and up to the corrections of the order
$u^{2}$, $w^{2}$, $uw$ and higher. To simplify the resulting expressions,
we have passed to the new parameters
\[ u \to u/128\pi^3, \quad w \to w/64\pi^3; \]
in (\ref{Zo}) and below they are denoted by the same symbols $u$ and $w$.

\section{RG functions and equations} \label{sec:RGE}

Let us briefly recall an elementary derivation of the RG equations;
detailed presentation can be found {\it e.g.} in \cite{Zinn,Book3}.
The RG equations are written for the renormalized Green functions
$W^{R} =\langle \Phi\cdots\Phi\rangle_{R}$, which differ from the
original (unrenormalized) ones $W =\langle
 \Phi\cdots\Phi\rangle$ only by
normalization (due to rescaling of the fields) and choice of
parameters, and therefore can equally be used for analyzing the
critical behaviour. The relation ${\cal S}_{R} (\Phi,e,\mu) =
{\cal S}(Z_{\Phi}\Phi,e_{0})$ between the bare (\ref{Action}) and
renormalized (\ref{ActionR}) action functionals results in the relations
\begin{equation}
W(e_{0},\dots) = Z_{\psi}^{N_{\psi}}
Z_{\psi^{\dagger}}^{N_{\psi^{\dagger}}} W^{R}(e,\mu,\dots).
\label{multi}
\end{equation}
between the Green functions. Here $N_{\psi}$ and
$N_{\psi^{\dagger}}$ are the numbers of corresponding fields
in the function $W$ (we recall that in our model $Z_{v}=1$);
$e_{0}=\{\lambda_{0}, \tau_{0}, u_{0}, w_{0}, a_{0}, m_{0},
\alpha_{0} \}$ is the full set of bare parameters and $e=\{
\lambda, \tau, u, w, a, m, \alpha  \}$ are their renormalized
analogs (we recall that $\alpha_{0}=\alpha$ and $m_{0}=m$);
the ellipsis stands for the other arguments (coordinates/momenta
or times/frequencies).

We use $\widetilde{\cal D}_{\mu}$ to denote the differential
operation $\mu\partial_{\mu}$ for fixed $e_{0}$ and operate on
both sides of the relation (\ref{multi}) with it. This gives the
basic RG differential equation:
\begin{equation}
\left\{ {\cal D}_{RG} + N_{\psi} \gamma_{\psi} + N_{\psi^{\dag}}
\gamma_{\psi^{\dag}} \right\} \,W^{R}(e,\mu,\dots) = 0,
\label{RG1}
\end{equation}
where ${\cal D}_{RG}$ is the operation $\widetilde{\cal D}_{\mu}$
expressed in the renormalized variables:
\begin{equation}
{\cal D}_{RG}\equiv {\cal D}_{\mu} + \beta_{u}\partial_{u} +
\beta_{w}\partial_{w}  + \beta_{a}\partial_{a} -
\gamma_{\lambda}{\cal D}_{\lambda} - \gamma_{\tau}{\cal D}_{\tau}
\label{RG2}
\end{equation}
and we have written ${\cal D}_{x}\equiv x\partial_{x}$ for any
variable $x$. The anomalous dimensions $\gamma$ are defined as
\begin{equation}
\gamma_{F}\equiv \widetilde{\cal D}_{\mu} \ln Z_{F} \quad {\rm for\ any\ quantity} \ F,
\label{RGF1}
\end{equation}
while the $\beta$ functions for the coupling constants $u$, $w$ and $a$ are
\begin{eqnarray}
\beta _{u} \equiv \widetilde {\cal D}_{\mu} u = u\,
(-\varepsilon-\gamma_{u}),
\nonumber \\
\beta_{w} \equiv \widetilde {\cal D}_{\mu} w =
w\,(-\xi-\gamma_{w}),
\nonumber \\
\beta_{a} \equiv \widetilde {\cal D}_{\mu} a = -a\gamma_{a},
\label{betagw}
\end{eqnarray}
where the second equalities come from the definitions and the
relations (\ref{Multy}). In principle, the dimensionless parameter
$\alpha$ should be treated as the fourth coupling constant, but the
corresponding $\beta$ function
\begin{eqnarray}
\beta_{\alpha}=\widetilde{\cal
D}_{\mu}\alpha=-\alpha\gamma_{\alpha} \label{Bal}
\end{eqnarray}
vanishes identically due to (\ref{exi}) and does not appear in the
subsequent relations.

The anomalous dimension corresponding to a given renormalization
constant $Z_{F}$ is found from the relation
\begin{equation}
\gamma_{F} = \left( \beta_{u}\partial_{u}+\beta_{w}\partial_{w}
+ \beta_{a}\partial_{a} \right) \ln Z_{F} \simeq  -
\left(\varepsilon {\cal D}_{u}+\xi {\cal D}  _{w}\right) \ln Z_{F}.
\label{GfZ}
\end{equation}
The first equality follows from the definition (\ref{RGF1}),
expression (\ref{RG2}) for the operation $\widetilde{\cal D}_{\mu}$
in renormalized
variables, and the fact that the $Z$'s depend only on the
completely dimensionless coupling constants $u$, $w$ and $a$. In
the second (approximate) equality, we only retained the
leading-order terms in the $\beta$ functions (\ref{betagw}), which
is sufficient for the first-order approximation. The leading-order
expressions (\ref{Zo}) for the renormalization constants have the
forms
\begin{equation}
Z_{F} = 1 + \frac{u}{\varepsilon}{\cal A}_{F}(a,\alpha) + \frac{w}{\xi}
{\cal B_{F}}(a,\alpha).
\label{Zf}
\end{equation}
The factors $\varepsilon$ and $\xi$
in (\ref{GfZ}) cancel the corresponding poles contained in the expressions
(\ref{Zf}) for the constants $Z_{F}$, which leads to the final UV finite
expressions for the anomalous dimensions:
\begin{equation}
\gamma_{F} = - u {\cal A}_{F}(a,\alpha) - w {\cal B}_{F}(a,\alpha)
\label{gift}
\end{equation}
for any constant $Z_{F}$. Then equations (\ref{Zo}) give
\begin{eqnarray}
\gamma_{1} &=& R_{1}u/2 , \quad \gamma_{2} = R_{1}u/3 + w
(5+\alpha)/6, \quad
\gamma_{3} = 2R_{1}u ,
\nonumber \\
\gamma_{4} &=& 2R_{2}u + w \alpha a^{2}, \quad \gamma_{5} = uR_{1}/2
+ w \alpha (a-1)^{2},
\nonumber \\
\gamma_{6}  &=& uR_{1}(4a-1 )/2a.
\label{anom}
\end{eqnarray}

The multiplicative relations (\ref{ZZ}) between the
renormalization constants result in the linear relations between
the corresponding anomalous dimensions:
\begin{eqnarray}
\gamma_{\lambda} &=&  \gamma_{2} -\gamma_{1}, \quad
\gamma_{\tau} =  \gamma_{3} -\gamma_{2} ,
\nonumber \\
\gamma_{a} &=&  \gamma_{6} -\gamma_{1}, \quad \gamma_{u} =
-\gamma_{1}- 3\gamma_{2} +2\gamma_{4} + \gamma_{5},
 \nonumber \\
2\gamma_{\psi} &=& \gamma_{1} +\gamma_{2} -\gamma_{5}, \quad
2\gamma_{\psi^{\dag}}= \gamma_{1} - \gamma_{2} + \gamma_{5},
\label{aesoC}
\end{eqnarray}
while the exact relations (\ref{RenD}) result in
\begin{eqnarray}
\gamma_{w} =-\gamma_{\lambda}, \quad
\gamma_{m} =\gamma_{\alpha} =\gamma_{v} = 0.
\label{exi}
\end{eqnarray}
Along with (\ref{anom}), those relations give the final first-order explicit
expressions for the anomalous dimensions of the fields and parameters:
\begin{eqnarray}
\gamma_{\lambda} &=& - \gamma_{w} = -R_{1}u/6 + w(5+\alpha)/6, \
\gamma_{\tau} = 5R_{1}u/3-w(5+\alpha)/6,
\nonumber \\
\gamma_{u} &=& -uR - w(5+\alpha)/2 +w\alpha f(a), \
\gamma_{a} = (3a-1)R_{1}u /2a,
\nonumber \\
2\gamma_{\psi} &=& R_{1}u/3 -w\left[ 5w(\alpha-1)/6 +\alpha a(a-2)\right],
\nonumber \\
2\gamma_{\psi^{\dag}} &=&
2R_{1}u/3 +w\left[ 5w(\alpha-1)/6 +\alpha a(a-2)\right],
\label{KK}
\end{eqnarray}
where we have denoted $R=R_1-4R_2$ and $f(a)=2a^{2}+(a-1)^{2}$.

\section{Attractors of the RG equations and scaling regimes} \label{sec:FPS}

It is well known that possible asymptotic regimes of a
renormalizable field theoretic model are determined by the
asymptotic behaviour of the system of ordinary differential
equations for the so-called invariant (running) coupling constants
\begin{eqnarray}
{\cal D}_s \bar g_{i}(s,g) = \beta_{i} (\bar g), \quad
\bar g_{i}(1,g) = g_{i},
\label{Odri}
\end{eqnarray}
where $s=k/\mu$, $g= \{g_{i}\}$ is the full
set of couplings and $\bar g_{i}(s,g)$ are the
corresponding invariant variables. As a rule, the IR ($s\to0$) and
UV ($s\to\infty$) behaviour of such system is determined by fixed
points $g_{i*}$. The coordinates of possible fixed points are
found from the requirement that all the $\beta$ functions vanish:
\begin{eqnarray}
\beta_{i} (g_{*}) =0.
\label{fp}
\end{eqnarray}
The type of a given fixed point is determined by the matrix
\begin{equation}
\Omega_{ij} = \partial\beta_{i}/\partial g_{j} |_{g=g^*}\, ,
\label{OmegaDef}
\end{equation}
which appears in the linearized version of the system (\ref{Odri}) near that
point. For IR attractive fixed points (which we are interested in here)
the matrix $\Omega$ is positive, {\it i.e.} the real parts of all
its eigenvalues are positive. In the case at hand, the fixed points for
the full set of couplings $u$, $w$, $a$, $\alpha$ are
determined by the equations
\begin{equation}
\beta_{u,w,a,\alpha} (u_{*},w_{*},a_{*},\alpha_{*}) = 0,
\label{points}
\end{equation}
with the $\beta$ functions defined in the previous section.
However, in our model the attractors of the system (\ref{Odri})
involve, in general, two-dimensional surfaces in the full
four-dimensional space of couplings. Indeed, the function
(\ref{Bal}) vanishes identically, and the equation
$\beta_{\alpha}=0$ imposes no restriction on the parameter $\alpha$.
Thus it is convenient to consider the attractors of the system
(\ref{Odri}) in the three-dimensional space $u$, $w$, $a$; their
coordinates, matrix (\ref{OmegaDef}) and the critical exponents
depend, in general, on the free parameter $\alpha$. Furthermore,
in this reduced space some attractors will be not simply fixed
points, but also lines of fixed points, parametrized by the coupling $a$.
In the following we will use the term ``fixed point'' for all those
attractors, bearing in mind that their coordinates can depend on $\alpha$
and, in general, on $a$.

The couplings $u_{*}$ and $w_{*}$ should be positive (by definition,
$u\propto g^2 >0$ and $w \propto D_0/\lambda > 0$), so that
the point is admissible from the physics viewpoints if it satisfies the
conditions
\begin{equation}
u_{*} > 0, \quad w_{*} > 0
\label{posit}
\end{equation}
and can be IR attractive ($\Omega>0$) for some values of the model
parameters (in fact, all the above inequalities can be non-strict).

In the one-loop approximation, the $\beta$ functions are found from the
definitions (\ref{betagw}) and the explicit expressions (\ref{exi}) and
(\ref{KK}):
\begin{eqnarray}
\beta_{u} &=& u \left[ -\varepsilon+ Ru + w(5+\alpha)/2 -w\alpha
f(a) \right],
\nonumber \\
\beta_{w} &=& w \left[ -\xi - R_1 u/6 + w(5+\alpha)/6 \right],
\nonumber \\
\beta_{a} &=& uR_1(1-3a)/2,
\label{beta}
\end{eqnarray}
with $R=R_1-4R_2$ and $f(a)=2a^{2}+(a-1)^{2}$.

The general pattern of the attractors of the system (\ref{beta}) is rather
complicated and depends qualitatively on the values of the parameters
$R_{1,2}$ and $\alpha$. There are four fixed points:

(I) The line of Gaussian (free) fixed points: $u_{*}=w_{*}=0$,
$a_{*}$ arbitrary.

(II) The point with $w_{*}=0$, corresponding to the pure Potts model
(turbulent advection is irrelevant).

(III) The line of fixed points with $u_{*}=0$ and  arbitrary $a_{*}$
corresponding to the passively advected scalar without self-interaction.

(IV) The most nontrivial fixed point with $u_{*}\ne0$, $w_{*}\ne0$,
corresponding to the novel scaling regime (universality class):
both the advection and the self-interaction are relevant.

Let us discuss these points in more detail.

\subsection{Fixed points with $u_{*}=0$.}  \label{Gauss}

For the Gaussian fixed point (point I) $u_{*}=w_{*}=0$, $a_{*}$ arbitrary.
The only nonzero off-diagonal element in $\Omega$ is $\Omega_{au}$.
Thus the matrix $\Omega$ is triangular and its eigenvalues coincide with
the diagonal elements: $\Omega_{u}=-\varepsilon$, $\Omega_{w}=-\xi$,
$\Omega_{a}=0$. Here and below $\Omega_{i} =\Omega_{ii}$ denote the
diagonal elements (no summation over $i$). Vanishing of the last
eigenvalue reflects the fact that the point is degenerate.

The coordinates of the passive scalar point (point III) are:
\begin{eqnarray}
u_{*}=0, \quad w_{*}= \frac{6\xi}{(5+\alpha)},
\quad a_{*} \ {\rm arbitrary}.
\label{3f}
\end{eqnarray}
Now $\Omega_{ua}=\Omega_{wa}=\Omega_{uw}=0$, so the matrix $\Omega$ is
block-triangular, and the eigenvalues are given by the diagonal elements:
\begin{eqnarray}
\Omega_{a}=0, \quad \Omega_{w}=\xi, \quad
\Omega_{u}= - \varepsilon +3\xi - f(a) \frac{6\alpha\xi}{(5+\alpha)}.
\label{omega3}
\end{eqnarray}
The condition $\Omega_{w}>0$ also gives $w_{*}>0$.

The function $f(a)=2a^{2}+(a-1)^{2}$ achieves the minimum value
$f(1/3)=2/3$ at $a=1/3$, and in
$\Omega_{u}$ we can write $f(a) = f(1/3) + [f(a)-f(1/3)]$. This gives
\begin{eqnarray}
\Omega_{u} = \Omega_{u}|_{a=1/3} -  \frac{6\alpha\xi}{(5+\alpha)}
[f(a)-f(1/3)] >0 .
\label{A}
\end{eqnarray}
The second term is negative, so (\ref{A}) can be satisfied only if
$\Omega_{u}|_{a=1/3}>0$. This gives
\begin{eqnarray}
-(5+\alpha) \varepsilon +(15-\alpha) \xi >0.
\label{b}
\end{eqnarray}
This is the domain in the $\xi$--$\varepsilon$ plane where the point
can be stable. Now (\ref{A}) gives the restriction for $a_{*}$:
\begin{eqnarray}
6\alpha\xi [f(a)-f(1/3)] <  -(5+\alpha) \varepsilon +(15-\alpha) \xi,
\label{V}
\end{eqnarray}
which gives
\begin{eqnarray}
(a_{*}-1/3)^{2} < \frac {-(5+\alpha) \varepsilon +
(15-\alpha) \xi}{18\alpha\xi}.
\label{a}
\end{eqnarray}

We conclude that the admissible fixed points of the type III
form an interval
on the line (\ref{3f}) specified by the inequality (\ref{a}).
The region of IR stability in the $\xi$--$\varepsilon$ plane
is given by the inequalities (\ref{b}) and $\xi>0$, so that
the condition $w_{*}>0$ is automatically satisfied.

For $\alpha=0$ the boundary defined by the inequality (\ref{b}) is
$\xi=\varepsilon/3$. When $\alpha$ increases, it rotates counter clockwise,
and for $\alpha\to\infty$ tends to $\xi=-\varepsilon$.

\subsection{Fixed points with $u_{*}\ne0$.} \label{nfp}

Now let us turn to the fixed points with $u\ne0$. Then the equation
$\beta_{a}=0$ readily gives $a=1/3$. Furthermore, one has
$\Omega_{ua}=\Omega_{wa}=0$, and the eigenvalue $\Omega_{a}= - 3R_{1}u/2$
decouples. For $u>0$, it can be positive only if $R_{1}<0$. Thus in the
following we assume $R_{1}<0$; the case $R_{1}>0$ requires special attention
and will be discussed later.

We put $a=1/3$ in $\beta_{u,w}$ and arrive at a closed system of two
$\beta$ functions for the two couplings $u,w$ of the form:
\begin{equation}
\beta_{u} = u [-\varepsilon+ Au + Bw], \quad \beta_{w} = w [-\xi+ Cu + Dw].
\label{Beta}
\end{equation}
For our set of $\beta$ functions (\ref{beta}) the coefficients in
(\ref{Beta}) are:
\begin{equation}
A=R, \ B=(15-\alpha)/6, \ C=-R_{1}/6>0, \ D=(5+\alpha)/6>0,
\label{ABC}
\end{equation}
but it is instructive to discuss it first in a general form with arbitrary
real coefficients $A$--$D$. Now we are interested only in the fixed points
with $u\ne0$; there are two such points: the Potts-type point with $w=0$
and the full-scale point with $w\ne0$.

Pure Potts point (point II). Here $w_{*}=0$, $u_{*} = \varepsilon/R$. Now
$\Omega_{wu}=0$ and the reduced $2\times2$ matrix $\Omega$ is triangular.
Then the point is IR stable for $\varepsilon>0$ (so that $R>0$ since we
require that $u_{*}>0$) and $-\xi+Cu>0$. The last relation gives
$\xi< - \varepsilon R_{1}  /6R$, because $A/C>0$.
Thus point II
\[ u_{*} = \varepsilon/R, \quad w_{*}=0, \quad a_{*}=1/3 \]
can be physical only if $R>0$, $R_{1}<0$ (and any $\alpha$) and is IR
stable in the region
\begin{equation}
 \varepsilon>0, \quad \xi< -R_{1} \varepsilon /6R.
\label{Tri}
\end{equation}

The coordinates of the full-scale fixed point (point IV) are
\begin{equation}
u_{*}= (D\varepsilon - B\xi )/\Delta, \quad
w_{*}= (A\xi - C\varepsilon )/\Delta,
\quad \Delta = AD-BC,
\label{2}
\end{equation}
while the reduced matrix $\Omega$ can be written in the form
\begin{eqnarray}
\Omega_{uu}=Au_{*}, \quad
\Omega_{uw}= Bu_{*}, \quad
\Omega_{wu}= Cw_{*}, \quad
\Omega_{ww}= Du_{*}.
\label{3}
\end{eqnarray}
It is useful not to substitute explicit expressions (\ref{2}) for a while.
For the positive $2\times2$ matrix $\Omega$ the eigenvalues can be real and
positive, or they can be complex conjugate with positive real parts.
Thus the necessary and sufficient condition for the IR stability
can be is given by the two inequalities:
\begin{equation}
\textrm{det}\, \Omega>0, \quad \textrm{tr} \,\Omega >0.
\label{5}
\end{equation}

From (\ref{3}) we obtain
\begin{equation}
\textrm{det}\,\Omega = u_*w_*\Delta >0,
\label{6}
\end{equation}
which along with (\ref{posit}) shows that this point can be admissible
only if $\Delta >0$. For $\textrm{tr}\,\Omega$ we obtain:
\begin{equation}
\textrm{tr}\, \Omega = Au_* + Dw_* >0.
\label{7}
\end{equation}
There are three possibilities:

(1) $A>0$, $D>0$. In this case the inequality (\ref{7}) is an automatic
corollary of (\ref{posit}).

(2) $A<0, D<0$. Then (\ref{7}) contradicts to (\ref{posit}) and this
point cannot admissible.

(3) The parameters $A$ and $D$ are opposite in sign.
For definiteness, we assume that $A<0$, $D>0$.
In this case  from (\ref{7}) one obtains:
\begin{equation}
w_* > -Au_*/D >0,
\label{8}
\end{equation}
where the last inequality follows from $A<0$ and $u_*>0$.
The second inequality
in (\ref{posit}) is implied by the (\ref{8}) and thus becomes superfluous.
The region where the fixed point is IR attractive and positive is given by
the two inequalities
\begin{equation}
u_*>0, \quad Au_*+Dw_* >0.
\label{9}
\end{equation}

We conclude that point IV can be admissible if $\Delta >0$ and
at least one of the two parameters $A$ and $D$ is positive. The region
of admissibility is determined by the inequalities (\ref{posit}) if $A$
and $D$ are both positive, and by the inequalities of the type (\ref{9})
if $A$ and $D$ have different signs.

\

Now we turn to our specific model with the $\beta$ functions (\ref{beta})
and the coefficients (\ref{ABC}). Then the coordinates of the fixed point
IV have the forms
\begin{equation}
u_{*} = \frac{[\varepsilon(5+\alpha)-\xi(15-\alpha)]}{6 \Delta},
\quad w_{*} = \frac{[\varepsilon R_1/6+R\xi]}{\Delta}, \quad
a_{*}=1/3,
\label{wu4}
\end{equation}
where
\begin{equation}
\Delta = AD-BC = \frac{5}{6}(R+R_{1}/2) + \frac{\alpha}{6}(R-R_{1}/6).
\label{Delta}
\end{equation}

This point can be physical only if $\Delta>0$. Since $D>0$, $A=R$ can be of
either sign; we also recall that $R_{1}<0$. There are four different cases:
\begin{eqnarray}
(i)\ R>-R_{1}/2,
\nonumber \\
(ii)\ -R_{1}/2 >R >0,
\nonumber \\
(iii)\ 0>R >R_{1}/6 ,
\nonumber \\
(iv)\ R<R_{1}/6,
\label{casesR}
\end{eqnarray}
which should be discussed separately.

\

Case ({\it i}). Then automatically $R>0$ and $R-R_{1}/6>0$, so that
$\Delta>0$ for all $\alpha$.
The conditions that point IV is IR attractive coincide with the
conditions (\ref{posit}) that its coordinates are positive.

\

Case ({\it ii}). Then $R-R_{1}/6>0$. Thus $\Delta<0$ for small
$\alpha$, but becomes positive for $\alpha>\alpha_{0}$, where
\begin{equation}
\alpha_{0} = -5 \frac{(R+R_{1}/2)} {(R-R_{1}/6)} >0.
\label{Alo}
\end{equation}
The conditions that the point is IR attractive are again (\ref{posit}).

\

Case ({\it iii}). Then $R-R_{1}/6>0$, and the point IV can be
admissible for $\alpha>\alpha_{0}$ with the same $\alpha_{0}$. Now $A<0$,
and the conditions that the point is physical are given by the inequalities
(\ref{9}).

\

Case ({\it iv}). Then $\Delta<0$ for all $\alpha$, and point IV cannot
be admissible.

\

Now let us write the positivity conditions (\ref{posit}) for the cases
({\it i}) and ({\it ii}) in a more explicit form with the aid of
expressions (\ref{2}) and (\ref{wu4}):
\begin{equation}
(A\xi - C\varepsilon )>0 , \quad  (D\varepsilon - B\xi )>0.
\label{Stab}
\end{equation}
Since $R=A>0$, the first inequality is $\xi> (C/A)\varepsilon$.
Since $B>0$ for $\alpha<15$, the second inequality is $\xi< (D/B)\varepsilon$.
It is also important that
\begin{equation}
\frac{C}{A}-\frac{D}{B}= \frac{-\Delta}{AB}<0
\label{St}
\end{equation}
so that
\[ {C}/{A}<{D}/{B}. \]
Also note that $C/A$ and $D/B$ are positive.

For $\alpha>15$ we have  $B<0$, and the second inequality becomes
$\xi> (D/B)\varepsilon$ with  $D/B<0$.

Thus for $\alpha<15$, the admissibility region is the sector in the
upper right quadrant in the $\varepsilon$--$\xi$ plane, bounded by the
ray $\xi= (C/A)\varepsilon$ from below and $\xi= (D/B)\varepsilon$ from above.
When $\alpha$ grows, the upper ray $\xi= (D/B)\varepsilon$ rotates counter
clockwise and moves to the upper-left quadrant.
For case ({\it i}) $\alpha$ changes from 0 to $\infty$ and the ray changes
from $\xi = \varepsilon/3$ to $\xi = - \varepsilon$ (exactly like the
boundary (\ref{b}) of point III).

For case ({\it ii}) $\alpha$ changes from $\alpha_{0}$ to $\infty$ and
the ray changes from $\xi = -\varepsilon R_{1}/6R$ to $\xi = - \varepsilon$.
For the following it is important that for case ({\it ii}) one has
$-R_{1}/6R>1/3$. Also note that
at $\alpha=\alpha_{0}$ the two boundaries for point IV coincide with
each other (this is also obvious from relation (\ref{St}), in which
$\Delta=0$ for $\alpha=\alpha_{0}$) and with the boundary (\ref{Tri})
of point II.  Also note that
\begin{equation}
\alpha_{0}-15 = \frac{-20R}{(R-R_{1}/6)} <0.
\label{ak}
\end{equation}

\

Let us turn to the case ({\it iii}). Now the inequality
 $u_{*}>0$ in (\ref{9}) becomes
\[ B\xi <  D \varepsilon. \]
Now from (\ref{ak}) we see that $\alpha_{0}-15>0$,
so that $B<0$ and we obtain
\begin{equation}
\xi > \varepsilon (D/B) \quad {\rm with}\  D/B<0.
\label{Stab1}
\end{equation}

The second condition $Au_{*}+Dw_{*}>0$ in (\ref{9}) is
\[ \xi A(D-B) > \varepsilon D(C-A) \]
where $A=R<0$ and $(D-B) = (\alpha-5)/3 >0$. The last inequality holds
because $\alpha>\alpha_{0}>5$:
\[ \alpha_{0}-5 = -10 \frac{(R+R_{1}/6)} {(R-R_{1}/6)}>0. \]
Thus the second inequality is
\begin{equation}
\xi < \varepsilon \frac{D(C-A)}{A(D-B)} ,
\label{Stab2}
\end{equation}
where
\begin{equation}
\frac{D(C-A)}{A(D-B)} =
\frac{(5+\alpha)(R+R_{1}/6)}{2R(5-\alpha)} <0.
\label{Bound2}
\end{equation}

It is also important that ${D}/{B} > {D(C-A)}/{A(D-B)}$, because
\begin{equation}
  \frac{D(C-A)}{A(D-B)} - \frac{D}{B} = \frac{-\Delta D}{AB(D-B)}<0.
\label{StM}
\end{equation}
Thus the both inequalities (\ref{Stab1}), (\ref{Stab2}) are satisfied
in a sector in the upper left quadrant; the lower bound is (\ref{Stab1})
and the upper bound is (\ref{Stab2}).

From (\ref{Bound2}) it follows that, when $\alpha$ changes from $\alpha_{0}$
to $\infty$, the coefficient $D(C-A)/A(D-B)$ changes from $-R_{1}/6R$ to
$-(R+R_{1}/6)/2R$. The coefficient $D/B= (5+\alpha)/(15-\alpha)$ changes
from $-R_{1}/6R$ to $-1$. Thus for $\alpha=\alpha_{0}$ the domain has zero
width (this is also obvious from the relation (\ref{StM}), in which
$\Delta=0$ for $\alpha=\alpha_{0}$), and when $\alpha$ increases
it is getting wider.

\subsection{General pattern of the fixed points.} \label{gpf}

Now we are in a position to describe
the general pattern of the admissibility regions of the
fixed points in the $\varepsilon$--$\xi$ plane. In the one-loop
approximations, they all are sectors bounded by straight rays; in the
following, they are referred to as sectors I, II {\it etc}. There are
four different situations related to the four cases in (\ref{casesR}).

\

\begin{figure}
\begin{center}
\includegraphics[width=1.0\textwidth]{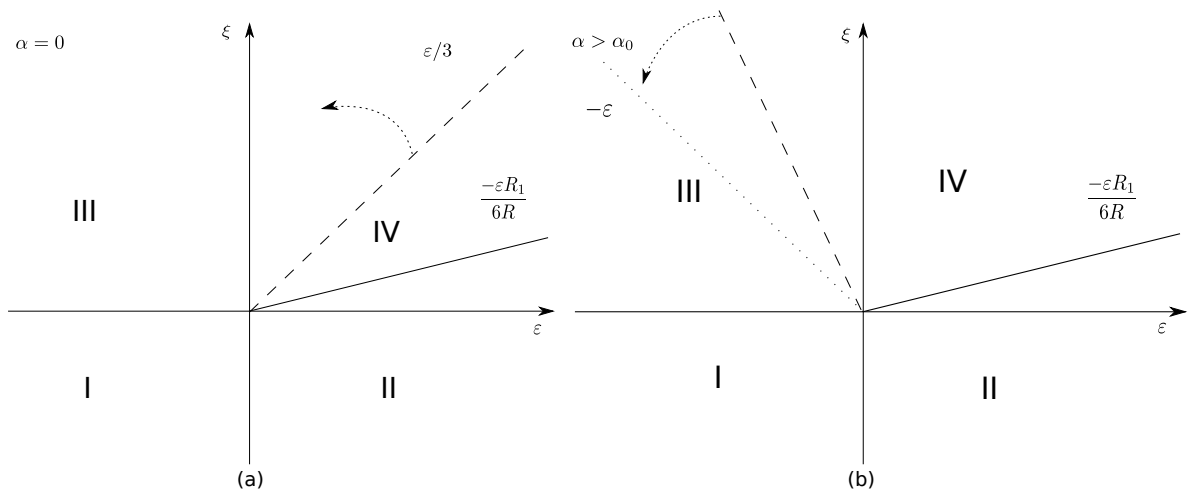}
\caption{Case ({\it i}): $\alpha=0$ (left) and $\alpha>\alpha_{0}$
(right).}
\label{fig2}
\end{center}
\end{figure}

Case ({\it i}) is illustrated by fig.~\ref{fig2}. The $\varepsilon$--$\xi$
plane is divided into four sectors I--IV without gaps or overlaps. The
boundary between sectors II and IV (solid line) is given by the ray
$\xi=-\varepsilon R_{1}/6R$. For $\alpha$ small, the boundary between
sectors III and IV (dashed line) lies in the upper right quadrant
(fig.~\ref{fig2}{\it a}). As $\alpha$ grows, it rotates counter clockwise
and for $\alpha>\alpha_{0}$ moves to the upper left quadrant
(fig.~\ref{fig2}{\it b}). Here and below, dotted lines denote the
$\alpha\to\infty$ limits of various boundaries. The boundary between
II and IV (solid line) is given by the ray $\xi=-\varepsilon R_{1}/6R$.

\

\begin{figure}
\begin{center}
\includegraphics[width=1.0\textwidth]{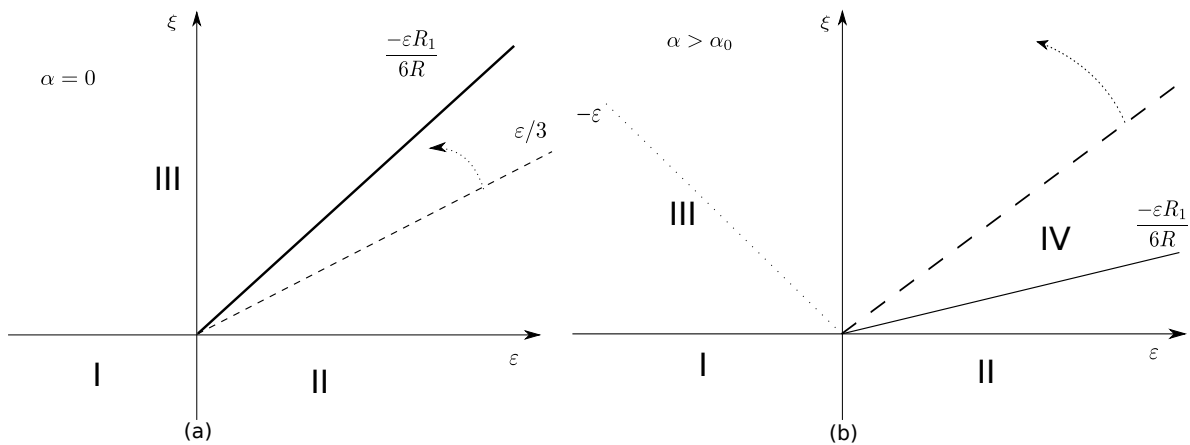}
\caption{Case ({\it ii}): $\alpha=0$ (left) and $\alpha>\alpha_{0}$
(right).}
\label{fig3}
\end{center}
\end{figure}

Case ({\it ii}) is illustrated by fig.~\ref{fig3}. For $\alpha$ small,
sector IV is absent, while sectors I--III cover the entire plane
$\varepsilon$--$\xi$ without gaps, but with an overlap between II and III
(fig.~\ref{fig3}{\it a}): the boundary $\xi=-\varepsilon R_{1}/6R$ of
sector II (solid line) lies above the boundary $\xi=\varepsilon/3$ of
sector III (dashed line). The existence of overlap means that, for the
corresponding values of $\varepsilon$ and $\xi$, the critical behaviour
is non-universal: it can be described by the fixed points II or III,
depending on the initial data for the problem (\ref{Odri}). As $\alpha$
grows, the boundary of sector III rotates, the overlap is getting thinner
and disappears for $\alpha=\alpha_{0}$. For $\alpha>\alpha_{0}$, there
is a gap between sectors II and III. Meanwhile, sector IV appears
and it fills that gap exactly. (fig.~\ref{fig3}{\it b}).

Thus starting with $\alpha=\alpha_{0}$, the $\varepsilon$--$\xi$ plane
is divided into four sectors I--IV with no gaps nor overlaps.

\

\begin{figure}
\begin{center}
\includegraphics[width=1.0\textwidth]{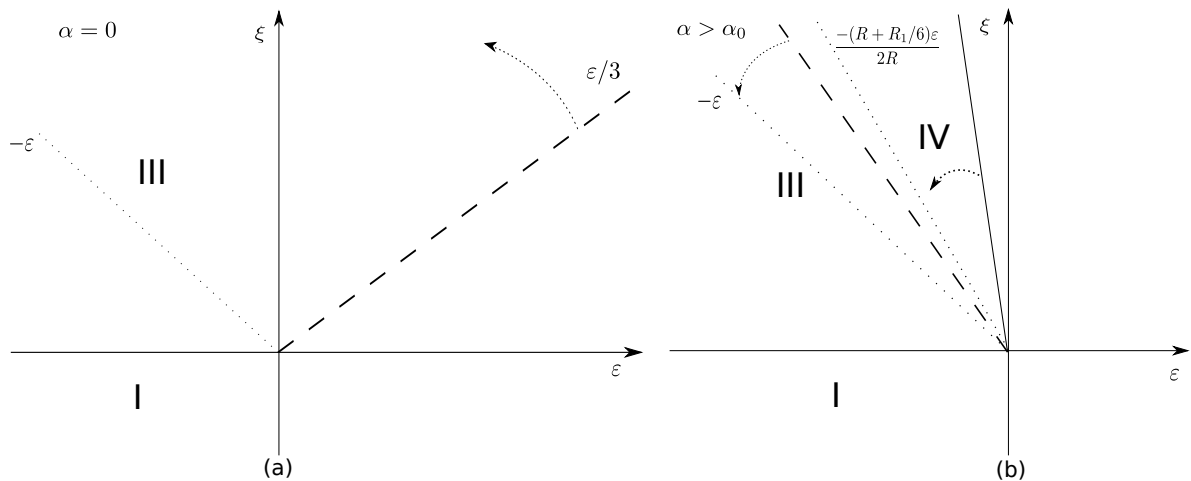}
\caption{Case ({\it iii}): $\alpha=0$ (left) and $\alpha>\alpha_{0}$
(right).}
\label{fig4}
\end{center}
\end{figure}

Case ({\it iii}). Sector II is absent, while IV is absent for small $\alpha$.
There is an empty space in the $\varepsilon$--$\xi$ plane, not covered
by any of the admissibility sectors (fig.~\ref{fig4}{\it a}). This is
interpreted as absence of a second-order phase transition for such values of
$\varepsilon$ and $\xi$. As $\alpha$ grows, the boundary of sector III
(dashed line) rotates counter clockwise, and at $\alpha=\alpha_{0}$ sector IV,
adjacent to III, appears in the upper left quadrant. Its right boundary
(solid line) also
rotates counter clockwise, but such that its width increases with $\alpha$.
There is no gap between III and IV for all $\alpha>\alpha_{0}$
(fig.~\ref{fig4}{\it b}).

Thus the turbulent mixing can lead to the emergence of a critical state
in a situation, where admissible fixed point does not exist for the original
static model (\ref{LG}) and for the equilibrium stochastic problem
(\ref{eq1}), (\ref{forceD}) without mixing.

\

\begin{figure}
\begin{center}
\includegraphics[width=0.6\textwidth]{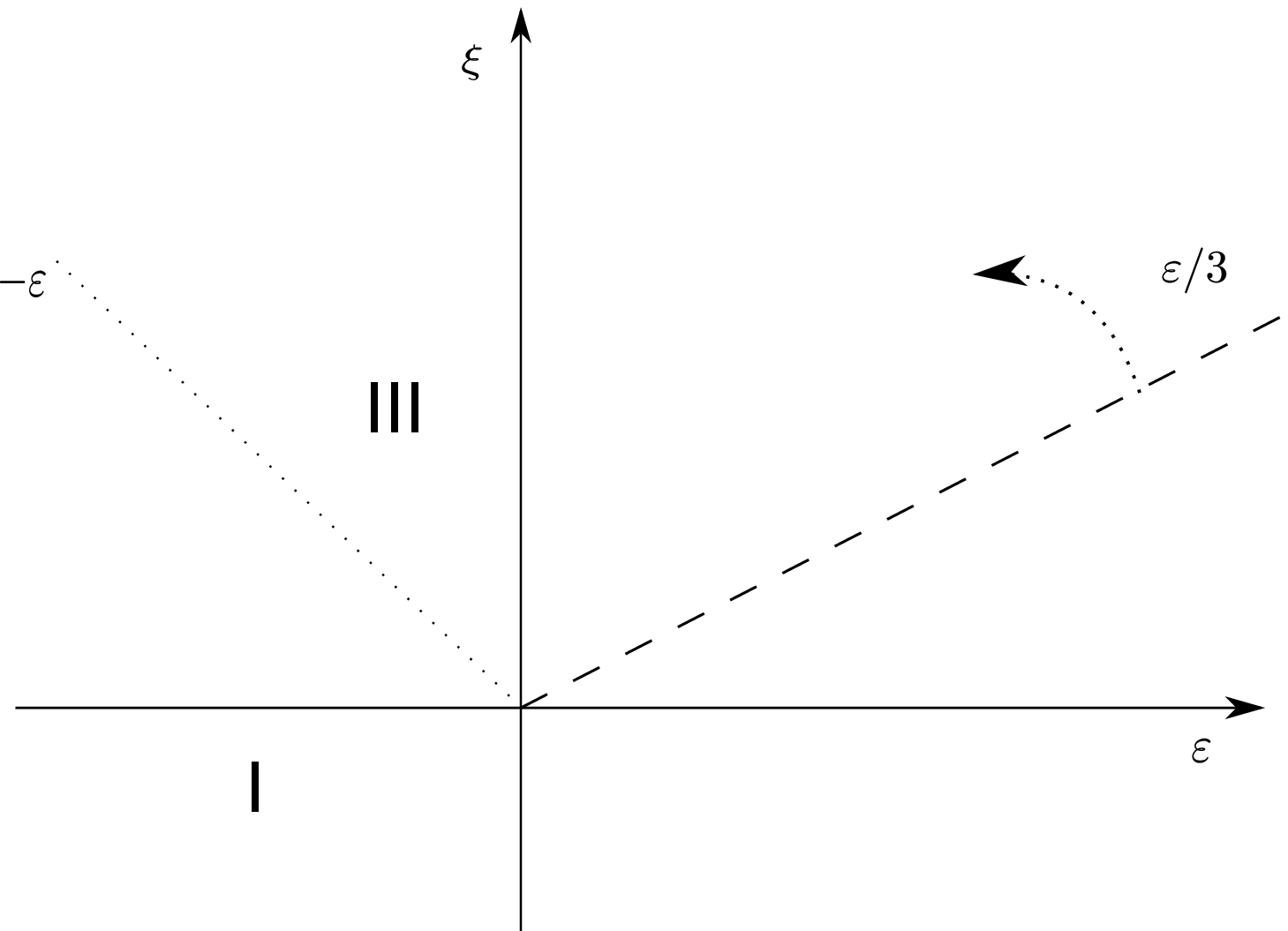}
\caption{Case ({\it iv}).}
\label{fig5}
\end{center}
\end{figure}

Case ({\it iv}). The most ``boring'' case, illustrated by fig.~\ref{fig5}.
Sectors II and IV are absent for all $\alpha$. When $\alpha$ grows, sector
III decreases, while the empty region grows.

\subsection{Fixed points with $R_{1}>0$.} \label{gR1}

It remains to discuss the case $R_{1}>0$ and $u_{*} \ne 0$. Then the fixed
points with $a_{*}=1/3$ cannot be IR attractive; see the discussion in
section~\ref{nfp}. However, nontrivial points for that case can be found
in terms of the new couplings $b=1/a$ and $v = wa^{2}$ with the $\beta$
functions
\begin{equation}
\beta_{b}= (-1/a^{2}) \beta_{a}, \quad
\beta_{v}= a^{2}\beta_{w}+ 2aw \beta_{a}.
\label{newB}
\end{equation}
Then
\begin{equation}
\beta_{b}= - uR_{1} b(b-3)/2.
\label{newBa}
\end{equation}
The relevant fixed point is $b_{*}=0$ with $\Omega_{bv}=\Omega_{bu}=0$, so
that $\Omega_{b}=3uR_{1}/2>0$ (for $u,R_{1}>0$) is an eigenvalue, and
$b$ decouples. We put $b=0$ in the other $\beta$ functions and again
obtain a closed system of the type (\ref{Beta}):
\begin{equation}
\beta_{u} = u[-\varepsilon+Ru-3\alpha v], \quad
\beta_{v} = v[-\xi - 19 uR_{1}/6].
\label{Bet2}
\end{equation}
We immediately see that $\Delta = - (19/2)\alpha R_{1} <0$, so that
the full-scale point with $u \ne 0$, $w \ne 0$ cannot be admissible.
The other possible fixed point is
\begin{equation}
v_{*}=0, \quad u_{*}= \varepsilon/R,
\label{b=0}
\end{equation}
with the IR stability conditions
\begin{equation}
 \varepsilon>0, \quad \xi< -  19R_{1} \varepsilon/6R,
\label{Bet3}
\end{equation}
so that $R>0$. Then the sector of admissibility lies in the lower right
quadrant. The point is similar to II in the sense that $v_{*}=0$ and the
advection is irrelevant.

\subsection{The plain Potts model} \label{ppm}

Let us conclude this section with a brief discussion of the original Potts
model with the hypertetrahedron symmetry. From (\ref{contract}) we obtain
$R=(n+1)^{2}(7-3n)$. The most interesting cases are $n=0$ (percolation
process in a moving medium) and $n=2$ (nematic-to-isotropic transition
in a liquid crystal). For $n=0$ we have $R_{1}=-1$, $R_{2}=-2$ and $R=7$;
those values belong to case ({\it i}) from (\ref{casesR}). For incompressible
or weakly compressible fluid, the most realistic values $\xi=4/3$ and $d=3$
correspond to the passive scalar regime (point III). As $\alpha$ increases,
the boundary between the regions III and IV moves, and for $\alpha$ large
enough the same values correspond to the new regime (point IV). Thus the
compressibility leads to the changeover in the type of critical behaviour
between two universality classes.

For $n=2$ we have $R_{2}=0$ and $R_{1}=R=9$. Thus the full-scale fixed point
cannot be admissible, while the regions of admissibility of the Potts-type
point II and the passive scalar case with $b_{*}=0$ lie in the lower right
quadrant (and thus are not very interesting for physics applications). For
small $\alpha$, the aforementioned physical values of $\xi$ and $\varepsilon$
belong to the passive scalar case with finite $a_{*}$ (point III), while for
$\alpha$ large enough they correspond to neither admissible point. Here the
growth of compressibility destroys the critical state.

For $n\ge3$ we have $R_{1}>0$ and $R<0$, so that the points with $b_{*}=0$
and the Potts-type point II cannot be admissible. Depending on the value of
$\alpha$, the physical values of $\xi$ and $\varepsilon$ belong to the
passive scalar case (\ref{b=0}) or lie in the ``desert'' in the
$\varepsilon$--$\xi$ plane with no admissible fixed points.


\section{Critical scaling and critical dimensions} \label{sec:DimeNS}

Existence of IR attractors in the RG equations implies existence of
asymptotic scaling regimes for all the Green functions in the IR range.
In dynamical models, critical dimensions $\Delta_{F}$ of the IR relevant
quantities (times/frequencies, coordinates/momenta, $\tau$ and the fields)
$F$ are given by the relations (see {\it e.g.} chap.~5 in \cite{Book3})
\begin{eqnarray}
\Delta_{F} = d^{k}_{F}+ \Delta_{\omega} d^{\omega}_{F} + \gamma_{F}^{*},
\quad  \Delta_{\omega}=2 -\gamma_{\lambda}^{*}.
\label{dim}
\end{eqnarray}
Here $d^{k,\omega}_{F}$ are the canonical dimensions of $F$, given in
table~\ref{table1}, and $\gamma_{F}^{*}$ is the value of the
corresponding anomalous dimension (\ref{RGF1}) at the given fixed point.
In our case, $\gamma_{F}^{*} = \gamma_{F} (u_{*},w_{*},a_{*},\alpha)$.
This gives:
\begin{eqnarray}
\Delta_{\psi}= (d-2)/2+\gamma^{*}_{\psi}, \quad
\Delta_{\psi^{\dag}}= (d+2)/2+\gamma^{*}_{\psi^{\dag}}, \quad
\Delta_{\tau} =2+\gamma^{*}_{\tau}.
\label{dim2}
\end{eqnarray}
The final results are obtained by substituting the coordinates of the
fixed points into the explicit one-loop expressions (\ref{KK}) for the
anomalous dimensions.

In particular, the response (Green) function in the IR range takes on the
asymptotic form (in the symmetric phase $\tau\ge0$)
\begin{eqnarray}
\langle \psi_{a} (t,{\bf x}) \psi^{\dag}_{b} (0,{\bf 0}) \rangle =
\delta_{ab}\, r^{-\Delta_{\psi}-\Delta_{\psi^{\dag}}} \,\, \Phi\left(
t r^{-\Delta_{\omega}}, \tau r^{\Delta_{\tau}} \right)
\label{Green}
\end{eqnarray}
with $r=|{\bf x}|$ and some scaling function $\Phi$.

For the Gaussian fixed point I the dimensions are trivial:
$\gamma^{*}_{F}=0$ for all $F$ (here and below, we present the results for
the anomalous dimensions, which are more graspable).

For point II the known one-loop results for the Potts model are recovered:
\begin{eqnarray}
\gamma_{\psi}^{*} = \frac{R_{1}\varepsilon}{6R},
\quad \gamma_{\psi^{\dag}}^{*} =  \frac{R_{1}\varepsilon}{3R}, \quad
\gamma_{\tau}^{*} = \frac{5R_{1}\varepsilon}{3R}, \quad
\gamma_{\lambda}^{*} = - \frac{R_{1}\varepsilon}{6R},
\label{dimPotts}
\end{eqnarray}
with corrections of order $O(\varepsilon^{2})$ and higher (no dependence
on $\xi$).

For the passive scalar point III one derives:
\begin{eqnarray}
\gamma_{\psi}^{*} = -\gamma_{\psi^{\dag}}^{*} =
\frac{(5-5\alpha-6\alpha a_{*}^{2} +12 \alpha a_{*})}{2(5+\alpha)}\,\xi ,
\quad
\gamma_{\lambda}^{*} = - \gamma_{\tau}^{*}  = \xi,
\label{dimPS}
\end{eqnarray}
where $a_{*}$ lies in the interval (\ref{a}).
The expressions for $\gamma_{\lambda,\tau}^{*}$, as well as the relation
$\gamma_{\psi}^{*}=-\gamma_{\psi^{\dag}}^{*}$
(so that $\Delta_{\psi}+\Delta_{\psi^{\dag}}=d$), are exact, because they
actually refer to the usual Kraichnan's model without self-interaction.

For the full-scale point IV we obtain:
\begin{eqnarray}
\gamma_{\psi^{\dag}}^{*} &=&
{(45R_{1}\varepsilon + 17R_{1}\alpha\varepsilon -180 R_{1}\xi
+12R_{1}\alpha\xi  -90R\xi+30R\alpha\xi)} / 216 \Delta \, ,
\nonumber\\
\gamma_{\psi}^{*} &=&
{(45R_{1}\varepsilon + R_{1}\alpha\varepsilon -90 R_{1}\xi
+6R_{1}\alpha\xi  +90R\xi-30R\alpha\xi)} / 216 \Delta \, ,
\nonumber\\
\gamma_{\tau}^{*} &=&
{(45R_{1}\varepsilon + 9R_{1}\alpha\varepsilon -150 R_{1}\xi
+10R_{1}\alpha\xi  -30R\xi-6R\alpha\xi)} / 36 \Delta
\label{dimFs}
\end{eqnarray}
with $\Delta$ from (\ref{Delta}) and the higher-order corrections in $\xi$
and $\varepsilon$. The exact result $\gamma_{\lambda}^{*} = \xi$ follows
from the first relation in (\ref{exi}) and the equation $\beta_{w}=0$ with
$w_{*}\ne0$.

\section{Conclusion} \label{sec:Conc}

We studied effects of turbulent mixing on the critical dynamics of a nearly
critical system, whose equilibrium behaviour is described by the
Ashkin-Teller-Potts model. The turbulent mixing was modelled by Kraichnan's
rapid-change ensemble: time-decorrelated Gaussian velocity field with the
power-like spectrum $\propto k^{-d-\xi}$. Special attention was paid to
compressibility of the fluid, because it leads to interesting qualitative
crossover phenomena.

The original stochastic problem was reformulated as a multiplicatively
renormalizable field theoretic model, which allowed us to apply the field
theoretic RG to the analysis of its IR behaviour. We showed that, depending
on the relation between the space dimension $d$, the exponent $\xi$ and
the degree of compressibility, the model reveals four types of possible IR
asymptotic behaviour,
associated with the four attractors (fixed points) of the RG equations.
Three fixed points correspond to known regimes: Gaussian (free) theory,
passively advected scalar field and the original Potts model without mixing.
The most interesting fixed point corresponds to a new type of critical
behaviour (universality class), where the self-interaction of the order
parameter and the turbulent mixing are equally important, and the critical
dimensions depend on $d$, $\xi$, the symmetry group ${\cal G}$ and the
compressibility parameter $\alpha$.

Explicit results were derived within the leading (one-loop) approximation,
that is, in the leading order of the double expansion in $\varepsilon$ and
$\xi$. Thus their validity for finite physical values of these parameters
can be called in question, especially because of large physical values
$\varepsilon=6-d=3$ and $\xi=4/3$. Careful discussion of this problem
requires analysis of higher-order corrections and applying some kind of
resummation procedure. Such analysis goes far beyond the scope of the present
paper, and we hope to address it in the future. Nevertheless, the discussion
of the RG flows, given in \cite{Anipoz} for a similar problem, suggests that
the pattern of critical regimes, obtained in the one-loop approximation,
appears robust with respect to higher-order corrections and can be preserved
for finite values of $\varepsilon$ and $\xi$.

Further investigation should account for conservation of the order parameter,
its feedback on the dynamics of the velocity statistics, finite correlation
time and non-Gaussian character of the advecting velocity field. This work
is in progress.

\section*{Acknowledgments}
The authors thank L\,Ts Adzhemyan, Michal Hnatich, Juha Honkonen and
M\,Yu Nalimov for discussions.
NVA thanks the Organizers of the International Meeting
``Conformal Invariance, Discrete Holomorphicity and Integrability''
(Helsinki, 10--16 June 2012) and the Department of Theoretical Physics
in the University of Helsinki for their kind hospitality.
The work was supported in part by the Russian Foundation for Fundamental
Research (grant No~12-02-00874a).


\section*{References}
\begin{thebibliography}{99}

\bibitem{Zinn} Zinn-Justin J 1989 {\it Quantum Field Theory and Critical
Phenomena} (Oxford: Clarendon)

\bibitem{Book3} Vasil'ev A N 2004 {\it The Field Theoretic Renormalization
Group in Critical Behavior Theory and Stochastic Dynamics}
(Boca Raton: Chapman \& Hall/CRC)

\bibitem{ATP} Ashkin J and Teller E 1943 {\it Phys. Rev.} {\bf 64} 178 \\
Potts R B 1952 {\it Proc. Camb. Phil. Soc.} {\bf 48} 106

\bibitem{Baxter} Baxter R J 1973 {\it J. Phys. C: Solid St. Phys.}
{\bf 6} L445

\bibitem{Golner} Golner G R 1973 {\it Phys. Rev.} A {\bf 8} 3419

\bibitem{Zia} Zia R K P and Wallace D J 1975 {\it J. Phys. A: Math. Gen.}
{\bf 8} 1495

\bibitem{Priest} Priest R G and Lubensky 1976 {\it Phys. Rev.} B {\bf 13}
4159; Erratum: B {\bf 14} 5125(E)

\bibitem{Amit} Amit D J 1976 {\it J. Phys. A: Math. Gen.} {\bf 9} 1441

\bibitem{Bonfim} de Alcantara Bonfim O F, Kirkham J E and  McKane A J
1980 {\it J. Phys. A: Math. Gen.} {\bf 13} L247 \\
de Alcantara Bonfim O F, Kirkham J E and  McKane A J
1981 {\it J. Phys. A: Math. Gen.} {\bf 14} 2391

\bibitem{VVV} Cardy J 2009 {\it J. Stat. Phys.} {\bf 137} 814 \\
Ikhlef Y and Cardy J 2009 {\it J. Phys. A: Math. Theor.} {\bf 42} 102001

\bibitem{CIDHI}
International Meeting {\it Conformal Invariance, Discrete Holomorphicity
and Integrability} (Helsinki, 10--16 June 2012).
Organizers Antti Kemppainen and Kalle Kyt\"{o}l\"{a},
https://wiki.helsinki.fi/display/mathphys/cidhi2012

\bibitem{Ivanov} Ivanov D Yu 2008 {\it Critical Behaviour of
Non-Ideal Systems} (Weinheim, Germany: Wiley-VCH)

\bibitem{Beysens} Beysens D, Gbadamassi M and Boyer L 1979
{\it Phys. Rev. Lett} {\bf 43} 1253 \\
Beysens D and Gbadamassi M 1979 {\it J. Phys. Lett.} {\bf 40} L565

\bibitem{Akira} Onuki A and Kawasaki K 1980 {\it Progr. Theor. Phys.}
{\bf 63} 122 \\
Onuki A, Yamazaki K and Kawasaki K 1981 {\it Ann. Phys.} {\bf 131} 217 \\
Imaeda T, Onuki A and Kawasaki K 1984 {\it Progr. Theor. Phys.} {\bf 71} 16

\bibitem{Ruiz} Ruiz R and Nelson D R 1981 {\it Phys. Rev.} A {\bf 23}
3224 {\bf 24} 2727 \\
Aronowitz A and and Nelson D R 1984 {\it Phys. Rev.} A {\bf 29} 2012

\bibitem{Satten} Satten G and Ronis D 1985 {\it Phys. Rev. Lett.}
{\bf 55} 91 \\
Satten G and Ronis D 1986 {\it Phys. Rev.} A {\bf 33} 3415

\bibitem{Chaotic} Lacasta  A M, Sancho J M and Sagu\'es F 1995
{\it Phys. Rev. Lett.} {\bf 75} 1791 \\
Berthier L 2001 {\it Phys. Rev.} E {\bf 63} 051503 \\
Berthier L, Barrat J-L and Kurchan J 2001 {\it Phys. Rev. Lett.}
{\bf 86} 2014 \\
Berti S, Boffetta G, Cencini M and Vulpiani A 2005
{\it Phys. Rev. Lett.} {\bf 95} 224501


\bibitem{Chan} Chan C K, Perrot F and Beysens D 1988 {\it Phys. Rev. Lett.}
{\bf 61} 412 \\
Chan C K, Perrot F and Beysens D 1989 {\it Europhys. Lett.} {\bf 9} 65 \\
Chan C K 1990 {\it Chinese J. Phys.} {\bf 28} 75 \\
Chan C K, Perrot F and Beysens D 1991 {\it Phys. Rev. A.} {\bf 43} 1826


\bibitem{AHH} Antonov N V, Hnatich M and Honkonen J 2006
{\it J. Phys. A: Math. Gen.} {\bf 39} 7867 \\
Antonov N V, Iglovikov V I and Kapustin A S 2009
{\it J. Phys. A: Math. Theor.} {\bf 42} 135001 \\
Antonov N V, Kapustin A S and Malyshev A V 2011
{\it Theor. Math. Phys.} {\bf 169} 1470

\bibitem{Alexa} Antonov N V and Ignatieva~A~A 2006
{\it J. Phys. A: Math. Gen.} {\bf 39} 13593 \\
Antonov N V, Ignatieva~A~A and Malyshev~A~V 2010
{\it Phys. Part. and Nuclei} {\bf 41} 998 \\
Antonov N V and Malyshev A V 2011
{\it Theor. Math. Phys.} {\bf 167} 444

\bibitem{AK} Antonov N V and Kapustin A S 2010
{\it J. Phys. A: Math. Theor.} {\bf 43} 405001

\bibitem{Anipoz} Antonov N V and Malyshev A V 2012
{\it J. Phys. A: Math. Theor.} {\bf 45} 255004

\bibitem{FGV} Falkovich G, Gaw\c{e}dzki K and Vergassola M 2001
{\it Rev. Mod. Phys.} {\bf 73} 913

\bibitem{JphysA} Antonov N V 2006 {\it J. Phys. A: Math. Gen.} {\bf 39}
7825



\end{thebibliography}
\end{document}